\documentclass{PoS}
\usepackage{amsmath}
\title{Cyclic Leibniz rule: a formulation of supersymmetry on lattice}

\ShortTitle{Cyclic Leibniz rule: a formulation of supersymmetry on lattice}

\author{Mitsuhiro Kato\\
 Institute of Physics, University of Tokyo, Komaba, Meguro-ku, Tokyo 153-8902, Japan \\
        E-mail: \email{kato@hep1.c.u-tokyo.ac.jp}}
\author{Makoto Sakamoto\\
        Department of Physics, Kobe University, Nada-ku, Hyogo 657-8501, Japan\\
        E-mail: \email{dragon@kobe-u.ac.jp}}

\author{\speaker{Hiroto So}%
         \thanks{
         This work was supported in part by
          Grants-in-Aid for Scientific Research (No.25400260 and No.25287049) 
 by the Japanese Ministry of Education, Science, Sports and Culture.}\\
    Department of Physics,  Ehime University, Bunkyou-chou 2-5, 
 Matsuyama 790-8577, Japan\\
        E-mail: \email{so@phys.sci.ehime-u.ac.jp}}
        
\abstract{
For the purpose of constructing supersymmetric(SUSY) 
theories on lattice, we propose a new type relation on lattice 
-cyclic Leibniz rule(CLR)- which is slightly different from an 
ordinary Leibniz rule. Actually, we find that  CLR can 
enlarge the number of SUSYs   and construct  more Nicolai mappings 
in a quantum-mechanical model. In this model, the exact mass degeneracy between fermion and 
boson is shown.
}

\FullConference{31st International Symposium on Lattice Field Theory LATTICE 2013\\
                 July 29 - August 3, 2013\\
                 Mainz, Germany}

\begin{document}

\section{Motivations}

One of purposes of constructing  lattice supersymmetry(SUSY) is to obtain  a method of  its nonperturbative analysis. 
Although the exact lattice supersymmetry is necessary for the analysis,  
there is an obstruction:~Leibniz rule on lattice theories with locality \cite{LR,bergner}.
The anticommutator between supercharges includes an infinitesimal  translation and 
its fact makes the realization of its lattice framework with locality  difficult. 
The realization of Leibniz rule  with locality on lattice  is  one of the difficulties.  
Recently, we found an alternative of the rule on lattice \cite{CLR}.
It is called  cyclic Leibniz rule(CLR).  The new rule is applicable to 
local lattice theories 
with a subset of full supersymmetry. This subset symmetry is nilpotent and  the algebra 
does not include translation operator.  To apply the symmetry for particle physics, 
the mass degeneracy between fermions and bosons, 
and the stability against quantum fluctuations such as the effective action 
are very important. We have to recognize that the subset symmetry keeps the degeneracy.
The stable property  is commented in the summary section.  
In this talk, we shall show the realization of the exact symmetry for a quantum mechanical model.  
In the model, we  explicitly show 
that fermion's and boson's masses are  exactly degenerate 
and  there exist  more conserved supercharges  and Nicolai mappings 
than the previous work's results \cite{NicolaiMap1,NicolaiMap2}.

\section{A key relation:cyclic Leibniz rule}
The setting of one-dimensional lattice systems is explained. 
The lattice size is denoted by $N_L$ and an integer $n$ is used as the lattice site labeling 
and the lattice constant $a$ is set to unity.   
Higher dimensional extensions are discussed in 
the last  section.

We write a general difference operator for a field $\phi_n$ as $(\Delta \phi)_n$ 
\begin{equation}
(\Delta \phi)_m \equiv \sum_n \Delta_{mn} \phi_n ,
\end{equation}
\noindent
and the symmetric field product rule on two fields $\phi_n$ and $\psi_n$ as $\{\phi,\psi \}^M_{\ell}$ 
\begin{equation}
\{\phi,\psi \}^M_{\ell} \equiv \sum_{mn}M_{\ell mn}\phi_m\psi_n =  \sum_{mn}M_{\ell nm}\phi_m\psi_n .
\end{equation}
\noindent
Introducing an inner product on a lattice space
\begin{equation}
(\phi,\psi) \equiv \sum_n \phi_n  \psi_n  ,
\end{equation}
\noindent
a usual Leibniz rule on lattice can be expressed as 
\begin{equation}
(\chi,\Delta\{\phi,\psi \}^M) = (\chi,\{\Delta\phi,\psi  \}^M)+(\chi,\{\phi,\Delta\psi   \}^M),  
\end{equation}
\noindent
for any three  bosonic fields, $\phi_n, \psi_n, \chi_n$.  If we take a generalized symmetric difference operator 
$\Delta=-\Delta^T$, the rule is expressed as 
\begin{equation}
(\Delta\chi,\{\phi,\psi \}^M) + (\chi,\{\Delta\phi,\psi  \}^M)+(\chi,\{\phi,\Delta\psi\}^M)=0.  
\end{equation}
 On the other hand, we present a new alternative relation on a set of a difference operator and a symmetric product rule; 
 that is cyclic Leibniz rule(CLR)  \cite{CLR}
\begin{equation}
(\Delta\chi,\{\phi,\psi\}^M) +(\Delta\phi,\{\psi,\chi\}^M)+ (\Delta\psi,\{\chi,\phi\}^M)=0 .
\label{CLR}
\end{equation}
\noindent
While  the Leibniz rule cannot be realized on lattice together with  locality and translational invariance, 
this  CLR can be realized locally.  Although there are many solutions for CLR,  
 we  here give a simple example
\begin{equation}
\Delta_{mn}=\frac{\delta_{m+1,n}-\delta_{m-1,n}}{2},~M_{\ell mn}=\frac{2\delta_{\ell+1,m}\delta_{\ell+1,n}
+ 2\delta_{\ell-1,m}\delta_{\ell-1,n}  + \delta_{\ell-1,m}\delta_{\ell+1,n}
+ \delta_{\ell+1,m}\delta_{\ell-1,n} }{6}.
\label{example}
\end{equation}
\noindent
In fact, This example (\ref{example})  is realized with locality and translational invariance.
 This CLR is sufficient to keep a subset of supersymmetry. 
If we extend to full supersymmetry, then we encounter the Leibniz rule and  it cannot be realized 
in the local lattice formulation.

\section{Supersymmetric complex quantum mechanics}
To construct $N=2$ supersymmetry  in one-dimensional lattice systems, 
we set the matter content of this model;  
for complex bosonic fields, $\phi_{+ n},F_{+ n}$, their complex conjugates, $\phi_{- n},F_{- n}$ and 
for fermionic fields, $\chi_{\pm n},\bar{\chi}_{\pm n}$. 
For these fields, 
$N=2$ supersymmetry transformation  
\begin{alignat}{4}
\delta_+ \phi_{+n} &= \bar{\chi}_{+n} ,   & \quad  \delta_- \phi_{+n} &=0 ,  &
\quad \bar{\delta}_+ \phi_{+n}&= 0 &\quad  \bar{\delta}_-\phi_{+n}&=  -{\chi}_{+n} \nonumber \\
\delta_+ \phi_{-n}     &=0    ,&  \quad {\delta}_-  \phi_{-n}&= \bar{\chi}_{-n} ,& 
\quad  \bar{\delta}_+ \phi_{-n}  &=-\chi_{-n}, &  \quad  \bar{\delta}_- \phi_{-n} &=  0,  \nonumber \\
\delta_+ \bar{\chi}_{+n}    &= 0 ,&  \quad \delta_- \bar{\chi}_{+n}  &= 0 ,&
 \quad \bar{\delta}_+ \bar{\chi}_{+n}&= -i(\Delta\phi_+)_n,&  
 \quad  \bar{\delta}_- \bar{\chi}_{+n} &= F_{+n} ,  \nonumber \\
 \delta_+ {\chi}_{+n}      &= F_{+n}    ,&  \quad  \delta_- \chi_{+n}&= i(\Delta\phi_+)_n ,&
 \quad \bar{\delta}_+ \bar{\chi_+}_n &=0 ,&  \quad  \bar{\delta}_- \chi_{+n}  &= 0 ,  \nonumber \\
  \delta_+   \bar{\chi}_{-n}              &= 0    ,&  \quad  \delta_- \bar{\chi}_{-n}&= 0 ,&
 \quad \bar{\delta}_+ \bar{\chi}_{-n} &= F_{-n} ,&  \quad  \bar{\delta}_-  \chi_{-n}&= -i(\Delta\phi_-)_n  ,  \nonumber \\
 \delta_+   \chi_{-n}               &=  i(\Delta\phi_-)_n   ,&  \quad \delta_-  \chi_{-n}&= F_{-n} ,&
 \quad \bar{\delta}_+ \chi_{-n}&= 0,&  \quad  \bar{\delta}_- \chi_{-n} &= 0 ,  \nonumber \\
  \delta_+    F_{+n}             &= 0    ,&  \quad  \delta_- F_{+n}&= -i(\Delta\bar{\chi}_+)_n ,&
 \quad \bar{\delta}_+ F_{+n}&= -i(\Delta\chi_+)_n ,&  \quad  \bar{\delta}_-  F_{+n} &=0  ,  \nonumber \\
 \delta_+     F_{-n}  &= -i(\Delta\bar{\chi}_-)_n    ,&  \quad  \delta_-  F_{-n} &=  0,&
 \quad \bar{\delta}_+ F_{-n}&= 0,&  \quad  \bar{\delta}_-  F_{-n} &= -i(\Delta\chi_-)_n  ,
\label{SUSY1}
\end{alignat}
\noindent
is defined. The algebra 
\begin{equation}
\{ \delta_{\pm},\delta_{\pm} \}=
\{ \bar{\delta}_{\pm},\bar{\delta}_{\pm} \} =\{ \delta_{\pm},\delta_{\mp} \}=
\{ \bar{\delta}_{\pm},\bar{\delta}_{\mp} \}=0 .
\end{equation}
\noindent
follows and  the remaining algebra 
\begin{equation}
\{\bar{\delta}_+,\delta_+\}+\{\bar{\delta}_-,\delta_-\} = -2i\Delta
\end{equation}
\noindent
is broken  by $O(a)$ since  this  $\Delta$ is dissatisfied with a Leibniz rule\cite{LR}. 
By using CLR, we shall  construct exactly  supersymmetric complex quantum mechanics on lattice. 
The supersymmetric action is 
\begin{eqnarray} 
S&=&(\Delta\phi_{-},\Delta\phi_+)- i(\Delta\bar{\chi}_-,\chi_+)-i(\Delta\bar{\chi}_+,\chi_-)+ (F_-,F_+)  \nonumber\\
&& +i(\phi_+,mF_+) +i(\bar{\chi}_+,m\chi_+)+i(\phi_-,m^*F_-)+i(\bar{\chi}_-,m^*\chi_-) \nonumber\\
&& +i\lambda (\{\phi_+,\phi_+\}^M,F_+)+ 2 i\lambda (\{\bar{\chi}_+,\phi_+\}^M,\chi_+) \nonumber\\
&&+ i\lambda^* (\{\phi_-,\phi_-\}^{\bar{M}},F_-)+ 2i \lambda^* (\{\bar{\chi}_-,\phi_-\}^{\bar{M}},\chi_-) ,
\label{action1}
\end{eqnarray}
\noindent
where the coefficient $\bar{M}_{\ell mn}$ of $\{,\}^{\bar{M}}$ is the complex conjugate of $M_{\ell mn}$. 
If we use a simple symmetric difference operator as $\Delta$, the doubling problem is generated. 
To avoid the problem, we can introduce a supersymmetric Wilson term by replacing a mass $m$  to $m\delta_{m,n}+H_{m,n}$, where 
\begin{equation}
H_{m,n}\equiv r\frac{\delta_{m+1,n}+\delta_{m-1,n}-2\delta_{m,n}}{2}
\end{equation}
 \noindent
with a Wilson parameter $r$.  
This action is  exactly invariant under the subset  transformation  
\begin{equation}
(\delta_{+},\delta_{-}) 
\label{SUSY2}
\end{equation}
\noindent
and its algebra is
\begin{equation}
\{\delta_+,\delta_-\}=\delta^2_{\pm}=0.
\end{equation}
Actually, under this transformation   (\ref{SUSY2}),  it is  shown that the action  (\ref{action1})  is invariant
\begin{eqnarray}
\delta_+S&=& i\lambda(\{\bar{\chi}_+,\phi_+\}^M  , F_+ )+ i\lambda(\{\phi_+,\bar{\chi}_+\}^M , F_+ )-2i\lambda (\{\bar{\chi}_+,\bar{\chi}_+\}^M,\chi_+ ) 
 \nonumber \\
&& -2i\lambda(\{\bar{\chi}_+,\phi_+\}^M,F_+) + i \lambda^*(\{\phi_-,\phi_-\}^{\bar{M}},-i\Delta\bar{\chi}_-)
 -2i  \lambda^* (\{\bar{\chi}_-,\phi_-\}^{\bar{M}},i\Delta\phi_-)=0, \nonumber \\
\delta_-S&=& 
i \lambda(\{\phi_+,\phi_+\}^{{M}},-i\Delta\bar{\chi}_+) -2i  \lambda (\{\bar{\chi}_+,\phi_+\}^{{M}},i\Delta\phi_+) \nonumber \\
&&+i\lambda^*(\{\bar{\chi}_-,\phi_-\}^{\bar{M}}  , F_- )+ i\lambda^*(\{\phi_-,\bar{\chi}_-\}^{\bar{M}} , F_- )
-2i\lambda^* (\{\bar{\chi}_-,\bar{\chi}_-\}^{\bar{M}},\chi_- ) \nonumber \\
&&-2i\lambda^*(\{\bar{\chi}_-,\phi_-\}^{\bar{M}},F_-)=0, 
\end{eqnarray}
\noindent
where we used CLR (\ref{CLR}) the symmetric property of the inner product  and the field product rule.


The subset (\ref{SUSY2}) of the full SUSY (\ref{SUSY1})  can be expressed by superfield formulation 
by introducing Grassmann numbers, $\theta_+,\theta_-$ with $\theta^2_{\pm}=0$. 
We define 
\begin{eqnarray}
\Phi_{+n} &\equiv& \phi_{+n}+\theta_+ \bar{\chi}_{+n} \nonumber \\
\Upsilon_{+n}& \equiv &   F_{+n}-i\theta_- (\Delta\bar{\chi}_+)_n \nonumber \\
\Psi_{+n} & \equiv &  \chi_{+n} +\theta_+F_{+n} + i\theta_- (\Delta\Phi_+)_{n}  \nonumber \\ 
S_{+n} &\equiv &   \bar{\chi}_{+n}  \nonumber \\
\Phi_{-n} &\equiv& \phi_{-n}+\theta_- \bar{\chi}_{-n} \nonumber \\
\Upsilon_{-n}& \equiv &   F_{-n}-i\theta_+ (\Delta\bar{\chi}_-)_n \nonumber \\
\Psi_{-n} & \equiv &  \chi_{-n} + \theta_+F_{-n} + i\theta_+ (\Delta\Phi_{-})_n \nonumber \\
S_{-n}  & \equiv  &    \bar{\chi}_{-n}  .
\label{SF}
\end{eqnarray}
\noindent
Combining  (\ref{SUSY2}) and (\ref{SF}), the transformation for the superfields can be written as 
\begin{alignat}{2}
\delta_+ \Phi_{+n}  &= S_{+n}  ,& \quad  \delta_- \Phi_{+n}&= 0, \nonumber \\
 \delta_+ \Phi_{-n}  &= 0 ,& \quad \delta_-\Phi_{-n} &= S_{-n}, \nonumber\\
\delta_+ \Psi_{+n}  &= \Upsilon_{+n} ,& \quad \delta_- \Psi_{+n} &=i (\Delta \Phi_+)_n ,  \nonumber \\
\delta_+ \Psi_{-n}  &=i (\Delta \Phi_-)_{n}  ,& \quad \delta_- \Psi_{-n} &= \Upsilon_{-n} ,  \nonumber  \\
\delta_+ \Upsilon_{+n}  &= 0 ,& \quad \delta_- \Upsilon_{+n} &=-i(\Delta S_+)_n , \nonumber \\
\delta_+ \Upsilon_{-n}  &=-i(\Delta S_-)_n  ,& \quad \delta_- \Upsilon_{-n} &= 0 , \nonumber \\
\delta_+ S_{+n}  &= 0  ,& \quad \delta_- S_{+n} &=  0,\nonumber \\
\delta_+ S_{-n}  &= 0   ,& \quad \delta_- S_{-n} &= 0 .
\label{SUSY3}
\end{alignat}
\noindent
The transformation (\ref{SUSY3}) implies 
\begin{eqnarray}
\delta_{\pm} F(\Phi_{\pm},\cdots)=\frac{\partial}{\partial \theta_{\pm}}F(\Phi_{\pm},\cdots)  ,
\end{eqnarray}
\noindent
where $F(\Phi_{\pm},\cdots)$ means arbitrary function of superfields (\ref{SF}). 
The action (\ref{action1}) is rewritten as 
\begin{eqnarray}
S&=&\int d^2\theta  (\Psi_-,\Psi_+) + i\int d^2\theta \theta_- (\Phi_+,m\Psi_+) 
+i\lambda\int d^2\theta  \theta_-(\{\Phi_+,\Phi_+\}^M,\Psi_+) \nonumber \\
&&+i \int d^2\theta \theta_- (\Phi_-,m^*\Psi_-)
+i \lambda^* \int d^2\theta  \theta_+(\{\Phi_-,\Phi_-\}^{\bar{M}},\Psi_-). 
\label{SFaction}
\end{eqnarray}
\noindent
where $\int d^2 \theta \theta_-\theta_+=1$ is used. 
We explicitly verify that the action (\ref{SFaction}) is invariant under (\ref{SUSY2}). 
Considering a Wilson term $H$ in $m$,  the following idenities and CLRs
\begin{equation}
(\Phi_{\pm},\Delta\Phi_{\pm})=(\Phi_{\pm},H\Delta\Phi_{\pm})=0 ,
\end{equation}
\noindent
\begin{equation}
(\{\Phi_+,\Phi_+\}^M,\Delta\Phi_+)=(\{\Phi_-,\Phi_-\}^{\bar{M}},\Delta\Phi_-)=0, 
\end{equation}
\noindent
the invariance of the action (\ref{SFaction})
\begin{eqnarray}
 \delta_+S&=& \int d^2\theta  \delta_+ (\Psi_-,\Psi_+)  - i\int d^2\theta \theta_- 
 (\delta_+ (\Phi_+,m\Psi_+)+  \lambda \delta_+(\{\Phi_+,\Phi_+\}^M,\Psi_+)   \nonumber \\
 && + i\int d^2\theta \theta_+  \Big( -(\Phi_-,im^*\Delta\Phi_-) - \lambda^*(\{\Phi_-,\Phi_-\}^{\bar{M}},i\Delta\Phi_-) \Big)  \nonumber \\
 &=& \int d^2\theta  \frac{\partial}{\partial\theta_+} (\Psi_-,\Psi_+)  - i\int d^2\theta \theta_- 
 \Big( \frac{\partial}{\partial\theta_+}  (\Phi_+,m\Psi_+)+  \lambda \frac{\partial}{\partial\theta_+} 
 (\{\Phi_+,\Phi_+\}^{{M}},\Psi_+) \Big)=0   \nonumber \\
  \delta_-S&=& \int d^2\theta  \delta_- (\Psi_-,\Psi_+)  - i\int d^2\theta \theta_+ 
 \Big(\delta_- (\Phi_-,m^*\Psi_-)+  \lambda^* \delta_-(\{\Phi_-,\Phi_-\}^{\bar{M}},\Psi_-)  \Big)  \nonumber \\
 && + i\int d^2\theta \theta_-  \Big( -(\Phi_+,im\Delta\Phi_+) - \lambda(\{\Phi_+,\Phi_+\}^{M},i\Delta\Phi_+ \Big)  \nonumber \\
 &=& \int d^2\theta  \frac{\partial}{\partial\theta_-} (\Psi_-,\Psi_+)  - i\int d^2\theta \theta_+ 
 \Big(\frac{\partial}{\partial\theta_-}  (\Phi_-,m^*\Psi_-)+  \lambda^* \frac{\partial}{\partial\theta_-} (\{\Phi_-,\Phi_-\}^{\bar{M}},\Psi_-)\Big) \nonumber \\
&& =0   
 \end{eqnarray}
\noindent
is shown.

\section{Supersymmetry, Mass degeneracy and Nicolai mappings}

In applying supersymmetry to particle physics,  it is necessary for the mass degeneracy between boson and fermion  
in the symmetry limit. The degeneracy  is clarified by using Ward-Takahashi relations 
by the subset supersymmetry (\ref{SUSY2}).  
From $\delta_+ \langle \phi_{+m} \chi_{-n}\rangle =0$ and  $\delta_- \langle \phi_{-m} \chi_{+n}\rangle =0$, 
the following two relations 
\begin{eqnarray}
\langle \phi_{+m}  (\Delta \phi_{-})_n \rangle & =&  i \langle  \bar{\chi}_{+m}\chi_{-n}  \rangle \nonumber \\
\langle \phi_{-m}  (\Delta \phi_{+})_n \rangle & =&  i \langle  \bar{\chi}_{-m}\chi_{+n}  \rangle 
\label{WT1}
\end{eqnarray}
\noindent
are derived.  We can calculate   boson's and fermion's masses($m_b,~m_f$) by taking the limit $\vert m-n \vert  \rightarrow \infty$ 
of  two-point functions,
\begin{equation}
\lim_{\vert n-m \vert \rightarrow \infty} \langle \phi_{+m}   \phi_{-n} \rangle =  K_b e^{-m_b\vert n-m \vert} 
\end{equation}
\noindent
and
\begin{equation}
\lim_{\vert n-m \vert \rightarrow \infty} \langle \bar{\chi}_{+m}   \chi_{-n} \rangle = K_f e^{-m_f\vert n-m \vert} ,
\end{equation}
\noindent
with $K_b$ and $K_f$ are constants. In the case of a symmetric difference operator, 
we can explicitly calculate 
\begin{equation}
\lim_{m-n\rightarrow \infty} \langle \phi_{+m}  (\Delta \phi_{-})_n \rangle  =-(\sinh m_b)
 K_be^{-m_b (m-n)} , 
\end{equation}
\begin{equation}
\lim_{m-n\rightarrow \infty} \langle \bar{\chi}_{+m}   \chi_{-n}  \rangle = K_f e^{-m_f (m-n)} .
\end{equation}
\noindent
Therefore,  from (\ref{WT1}) we can obtain  
\begin{equation}
-i(\sinh m_b) K_b=K_f,~m_b=m_f .
\end{equation}
\noindent
This result implies the exact mass degeneracy between fermion and boson.

Another result by our CLR is the existence of two kinds of local Nicolai mappings. 
In the existence, CLR plays a very important role as follows. 
After eliminating auxiliary fields $F_{\pm}$, 
the action is reduced to 
\begin{eqnarray}
S'&=&(\Delta\phi_{-},\Delta\phi_+)- i(\Delta\bar{\chi}_-,\chi_+)-i(\Delta\bar{\chi}_+,\chi_-)
+i(\bar{\chi}_+,m\chi_+) +i(\bar{\chi}_-,m^*\chi_-) \nonumber \\
 &&   +2 i\lambda (\{\bar{\chi}_+,\phi_+\}^M,\chi_+)+  2i \lambda^* (\{\bar{\chi}_-,\phi_-\}^{\bar{M}},\chi_-)    \nonumber \\
&& +( m^* \phi_-+ \lambda^*\{\phi_-,\phi_-\}^{\bar{M}}, m \phi_++ \lambda\{\phi_+,\phi_+\}^M   .
\label{action2}
\end{eqnarray}
For this action, we can define two Nicolai mappings 

\begin{eqnarray}
\xi_{+n}&\equiv &  (\Delta\phi_-)_n+i m \phi_{+n}+ i\lambda\{\phi_+,\phi_+\}_m^M  \nonumber \\
\xi_{-n} & \equiv & (\Delta\phi_+)_n+ im^* \phi_{-n}+ i\lambda^*\{\phi_-,\phi_-\}_n^{\bar{M}}
\end{eqnarray}
\noindent
and 
\begin{eqnarray}
\xi'_{+n}&\equiv &  -(\Delta\phi_-)_n+i m \phi_{+n}+i \lambda\{\phi_+,\phi_+\}_n^M  \nonumber \\
\xi'_{-n} & \equiv & -(\Delta\phi_+)+ im^* \phi_{-n}+ i\lambda^*\{\phi_-,\phi_-\}_n^{\bar{M}} .
\end{eqnarray}
\noindent
To write down the bosonic part $S'_b$ of $S'$ as 
\begin{equation}
S'_b= \sum_{n}\xi_{-n}\xi_{+n} = \sum_{n}\xi'_{-n}\xi'_{+n}, 
\end{equation}
\noindent
it is necessary for  vanishing  cross terms(surface terms) conditions 
\begin{equation}
(\Delta\phi_-,m^* \phi_-+ \lambda^*\{\phi_-,\phi_-\}^{\bar{M}})=(\Delta\phi_+,m \phi_++ \lambda\{\phi_+,\phi_+\}^M)=0 .
\end{equation}
\noindent
These conditions just correspond to  CLR and a symmetric difference operator for any $\phi_{\pm}$.  
The mappings  are useful in calculating quantum effects in this model. 
The fermion part of the action  in the model is written as 
\begin{equation}
S_{f}= \sum_{mn}
(\bar{\chi}_{+m},\bar{\chi}_{-m})
\left(
\begin{array}{cc}
 i m\delta_{mn} + 2i\lambda M_{nmk}\phi_{+k}& -i \Delta_{nm}   \\
 -i \Delta_{nm}&  im^*\delta_{mn} + 2i\lambda^* \bar{M}_{nmk}\phi_{-k}  
\end{array}
\right)
\left(
\begin{array}{c}
\chi_{+n}  \\
\chi_{-n}
\end{array}
\right)   .  
\end{equation}
\noindent
When these fermions are transformed as the following way
\begin{equation}
\left(
\begin{array}{c}
 \chi_{+n},  \\
 \chi_{-n},  \\
 \bar{\chi}_{+n},  \\
  \bar{\chi}_{-n}   
\end{array}
\right) 
\rightarrow
\left(
\begin{array}{c}
 \chi'_{+n},  \\
 \chi'_{-n},  \\
 \bar{\chi}'_{+n},  \\
  \bar{\chi}'_{-n}   
\end{array}
\right) 
= 
\left(
\begin{array}{c}
 \chi_{+n},  \\
 -\chi_{-n},  \\
 \bar{\chi}_{+n},  \\
 - \bar{\chi}_{-n}   
\end{array}
\right) ,  
\end{equation}
\noindent
the signs of the fermionic kinetic terms in the action (\ref{action2}) are  changed  and  
the fermion determinant is unchanged in any lattice size because  the Jacobian is 
$(-1)^{2N_L}$. 
The two descriptions $(\chi_+,\cdots)$ and $(\chi'_+,\cdots)$
of the fermion  corresponds to two Nicolai mappings. 


\section{Summary}

We have constructed an $N=2$ supersymmetric quantum mechanical model on lattice. 
Although the realization of the supersymmetry in the model is a subset of full SUSY, the Ward-Takahashi identities 
lead  to the excat degeneracy between fermion and boson.  The new type of superfield formalism can  describe 
 the action and the transformation  very concisely.  
Surprisingly, our model has a nonrenormalizaion theorem with respect to mass and interaction terms. 
The detail analysis will  appear   in our  forthcoming paper \cite{NRT}.

To extend to higher-dimensional theories, we must reconsider CLR in higher dimensions carefully. 
As discussed in Ref.\cite{CLR}, it is difficult to maintain locality  in a  higher-dimensional extension of the CLR.

\end{document}